%% file: main.tex
\documentclass{article}
\usepackage{spconf,graphicx}
\usepackage{amsmath,amssymb}
\usepackage{hyperref}
\usepackage{booktabs, enumitem, setspace}
\usepackage{capt-of}
\usepackage{etoolbox}

\usepackage{xcolor}

\newif\ifdraftnotes
\draftnotestrue

\definecolor{krcolor}{RGB}{35,100,190}
\definecolor{yjcolor}{RGB}{155,65,165}
\definecolor{todocolor}{RGB}{200,85,20}

\newif\ifmk
\mktrue

\newif\ifmk
\mktrue

\newcommand{\ours}{FOAdapter}

\hypersetup{hidelinks}

\title{FOAdapter: A Plug-and-Play Spatial Tokenizer\\for Low-Bitrate FOA Speech Coding}
\name{
Kirak Kim$^{1}$\sthanks{These authors contributed equally to this work.} \qquad
Yoonjeong Park$^{2*}$ \qquad
Juhan Nam$^{1,2}$ \qquad
Sungyoung Kim$^{1}$ \qquad
Minje Kim$^{3}$
}

\address{
$^{1}$ Graduate School of Culture Technology, KAIST, Daejeon, Republic of Korea \\
$^{2}$ Kim Jaechul Graduate School of AI, KAIST, Daejeon, Republic of Korea \\
$^{3}$ Siebel School of Computing and Data Science, University of Illinois Urbana-Champaign, USA
}

\begin{document}
\ninept
\maketitle
\begin{abstract}
Spatially encoded speech conveys not only linguistic information but also spatial cues. This enables listeners to distinguish and localize speakers, contributing to more natural and realistic auditory experiences in applications such as remote conferencing and immersive media. Neural speech codecs (NSCs) enable high-quality reconstruction of monaural speech at increasingly low bitrates while preserving linguistic information, but their ability to effectively preserve spatial cues remains limited. We propose \ours{}, a plug-and-play spatial tokenizer that leverages a pretrained monaural speech codec to efficiently represent both linguistic and spatial information in first-order ambisonic (FOA) speech at low bitrates. Comprehensive evaluation demonstrates that \ours{} consistently preserves spatial information across all evaluated codecs. 
Source code and audio examples are available on our demo website\footnote{\href{https://kirak-kim.github.io/foadapter-demo/}{https://kirak-kim.github.io/foadapter-demo/}}.
\end{abstract}
\input{sec/1_intro}
\input{sec/2_methodology}
\input{sec/3_experimental_setup}
\input{sec/4_experimental_results}
\input{sec/5_conclusion}

\bibliographystyle{IEEEbib}
\bibliography{refs}

\end{document}

%% file: sec/1_intro.tex
\section{Introduction}
\label{sec:introduction}

\suppressfloats[t] 

\begin{figure}[t] 
    \centering
    \includegraphics[width=\linewidth]
        {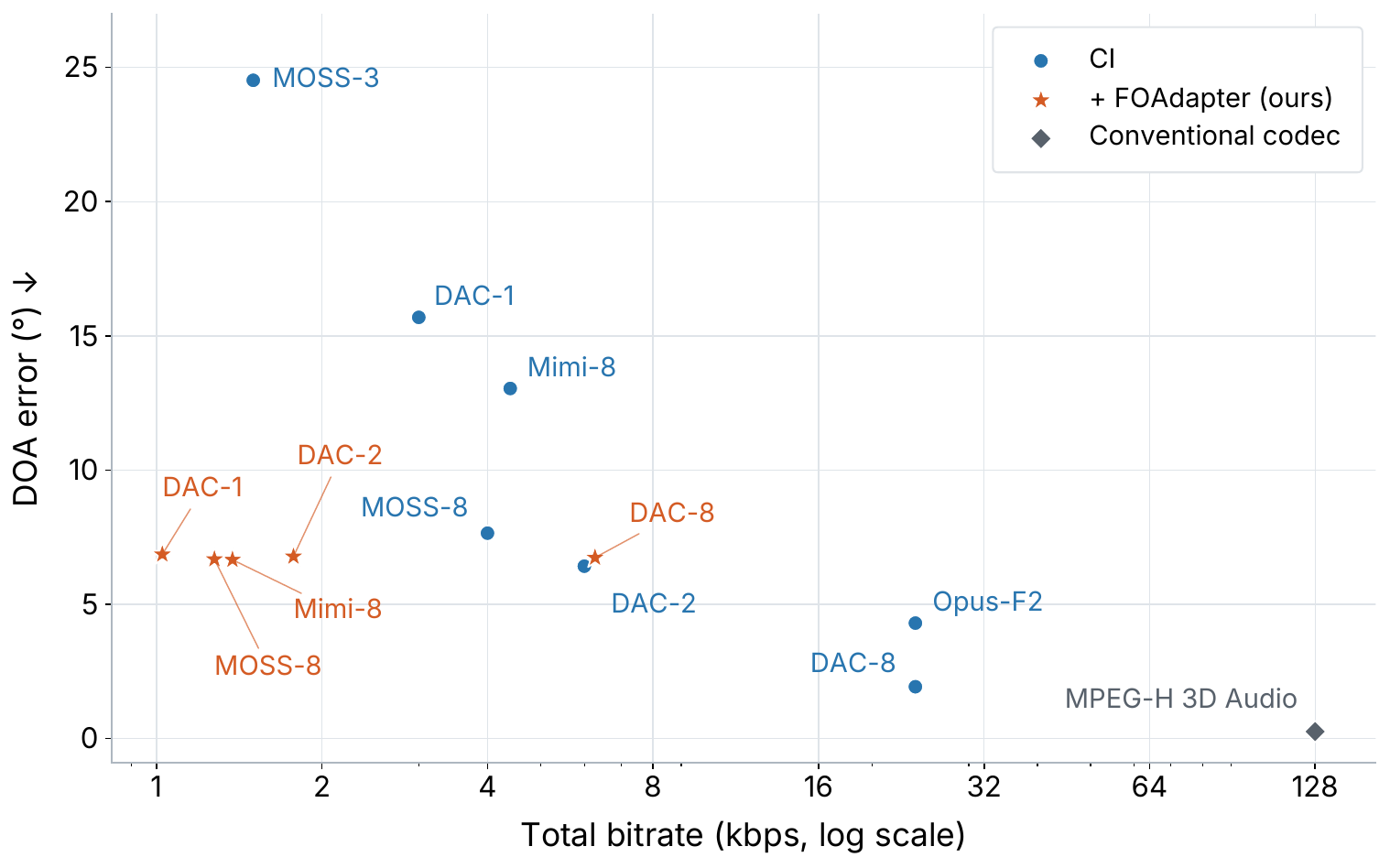}
    \caption{DOA error of reconstructed FOA signal versus total bitrate on the Spatial LibriSpeech
    test set. \ours{} achieves approximately $7^\circ$
    error across the evaluated monaural codecs using a 275-bit/s
    spatial stream.}
    \label{fig:bitrate_doa}
\end{figure}

\begin{figure*}[t]
    \centering
    \includegraphics[width=\textwidth]{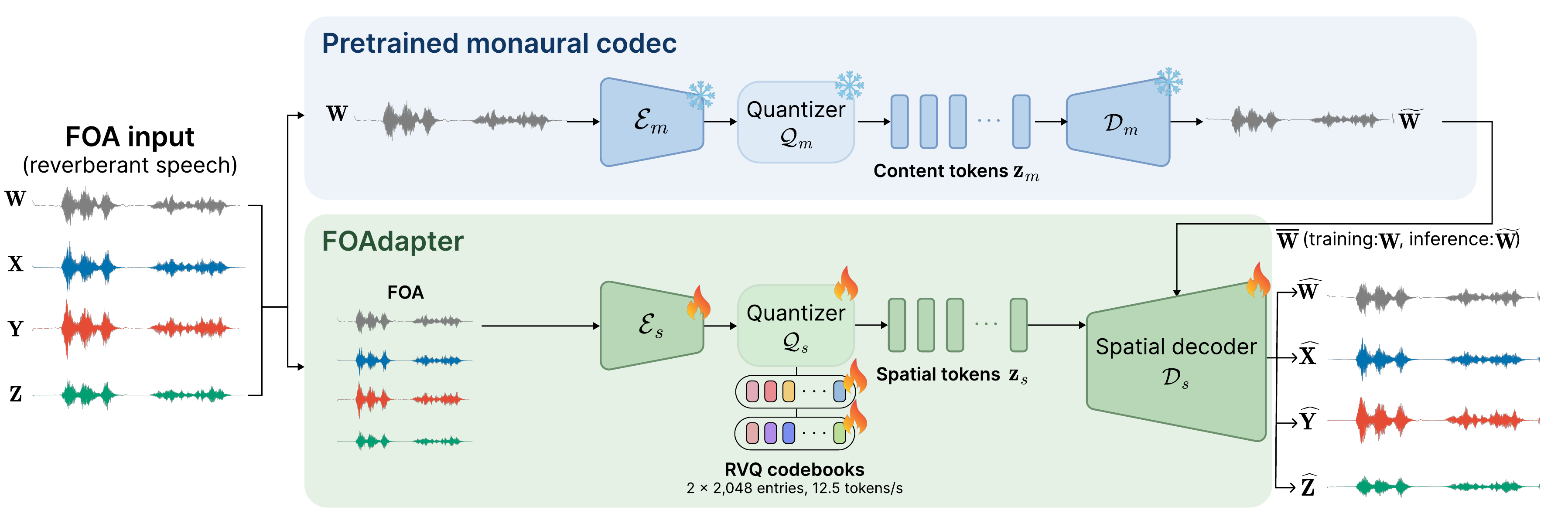}
    \caption{Overview of the proposed \ours{} framework.} 
    \label{fig:overview}
\end{figure*}

Spatial audio plays an essential role in immersive auditory experiences in communication and media applications. Its representation and delivery are supported by standardized coding systems such as MPEG-H 3D Audio~\cite{herre2014mpegh} and AC-4~\cite{etsi_ac4}, commercial technologies such as Dolby Atmos~\cite{dolby_atmos} and DTS:X~\cite{dtsx}, and codec-agnostic formats such as the Alliance for Open Media (AOMedia) Immersive Audio Model and Formats (IAMF)~\cite{aom_iamf}.

In this work, we are interested in multichannel single-speaker speech represented in first-order ambisonics (FOA). Such signals convey linguistic and acoustic information together with spatial cues that enable listeners to localize the speaker in three-dimensional space. FOA provides a compact representation of spatial audio using four spherical harmonic components: an omnidirectional component, \(W\), and three directional components, \(X\), \(Y\), and \(Z\)~\cite{zotter2019ambisonics}. By compactly encoding the directional structure of the sound field, this representation enables efficient manipulation of spatial information, including rotation of the sound scene and binaural rendering that adapts to the listener's orientation.

Neural speech codecs (NSCs) have enabled high-quality reconstruction of speech at increasingly low bitrates while preserving linguistic information ~\cite{wavtokenizer,mosstokenizer, yi2026hallucination}. Beyond waveform compression, the discrete speech tokens produced by such low-bitrate NSCs are widely used as representations for speech generation tasks ~\cite{speechtokenizer, valle, audiolm, moshi}. Advances in quantization~\cite{wavtokenizer, languagecodec}, semantic supervision~\cite{yi2026hallucination, speechtokenizer}, and large-scale pretraining~\cite{mosstokenizer} have further improved the acoustic and semantic information captured by these tokens.
However, most speech tokenizers are designed for monaural (mono) or stereo speech and do not explicitly encode the multichannel spatial structure represented by formats such as FOA. 

A straightforward way to extend a mono tokenizer to FOA is to encode each channel independently. At the same per-channel bitrate, this quadruples the total bitrate without exploiting information shared across channels.
Moreover, lower bitrates in this channel-independent setting can compromise spatial-cue preservation~\cite{spatial_fidelity}.
Efficient neural coding of spatial audio has recently attracted increasing research interest. 
SpatialCodec~\cite{spatialcodec} combines a pretrained mono codec for the reference channel with a spatial branch that encodes relative spatial information and predicts filters to reconstruct the remaining channels. This design leverages the speech coding capabilities of the pretrained mono codec while efficiently representing inter-channel spatial relationships. However, it targets eight-channel microphone-array recordings and was evaluated at a relatively high total bitrate of 12 kbps.
More recently, FOA Tokenizer~\cite{foatokenizer} extends WavTokenizer~\cite{wavtokenizer} to jointly encode four-channel FOA signals at 0.9 kbps using a spatial consistency objective.
Although it achieves efficient FOA coding at low bitrate, its reported word error rate of 67\% on their internal
SpatialVCTK dataset suggests room for improving speech-content preservation in that setting.
These studies motivate the following research question: \emph{Can we leverage a pretrained mono tokenizer for FOA speech coding at low bitrates while preserving both speech content and spatial information?}

In this work, we propose \ours{}, a causal spatial adapter that leverages a frozen pretrained mono speech tokenizer for FOA speech coding. 
\ours{} reconstructs the full FOA signal from the decoded $\mathbf{W}$ channel and a separate 275-bit/s spatial token stream. We target low-bitrate coding at a total bitrate of approximately 1.5 kbps, with a compact discrete representation intended to support generative speech  models, including speech language models.
%
Spatially guided contrastive learning encourages the spatial tokens to capture spatial cues while reducing their dependence on the speech content in $\mathbf{W}$.
Following the evaluation settings of prior work~\cite{spatialcodec, foatokenizer},
we evaluate single-talker speech with fixed source and receiver positions. 
Across the evaluated pretrained tokenizers, \ours{}
achieves consistent DOA errors of approximately $7^\circ$, with a trade-off in speech reconstruction quality. Token-swapping experiments further suggest that the spatial token stream conveys source direction largely independently of speech content.
A MUSHRA-inspired listening study further shows higher perceived speech quality than a comparable-bitrate channel-independent baseline.

%% file: sec/2_methodology.tex
\section{Methodology}
\label{sec:methodology}

\subsection{Adapting a pretrained mono codec to FOA coding}
Given a four-channel FOA signal of length $T$, $\mathbf{S}=[\mathbf{W},\mathbf{X},\mathbf{Y},\mathbf{Z}]^\top\in\mathbb{R}^{4\times T}$, we aim to reconstruct all four channels while preserving both speech content and spatial cues under a compact discrete representation. 

We extend a frozen pretrained mono codec to FOA speech coding with a separate low-rate spatial token stream. The frozen mono codec encodes the $\mathbf{W}$ channel, while a trainable spatial adapter, \ours{}, extracts spatial tokens from the full FOA signal to support reconstruction conditioned on the decoded $\mathbf{W}$.
The frozen monaural codec comprises an encoder $\mathcal{E}_m$, a quantizer $\mathcal{Q}_m$, and a decoder $\mathcal{D}_m$. \ours{} comprises a causal convolutional encoder $\mathcal{E}_s$ and decoder $\mathcal{D}_s$, together with a quantizer $\mathcal{Q}_s$. The mono token stream $\mathbf{z}_m$, spatial token stream $\mathbf{z}_s$, and reconstructed FOA signal $\widehat{\mathbf{S}}$ are given by
\begin{equation}
\label{eq:foa_reconstruction}
\begin{aligned}
\mathbf z_m &= \mathcal{Q}_m(\mathcal{E}_m(\mathbf W)), \quad
\widetilde {\mathbf W} = \mathcal{D}_m(\mathbf z_m),\\
\mathbf z_s &= \mathcal{Q}_s(\mathcal{E}_s(\mathbf S)), 
\quad\widehat{\mathbf S} = \mathcal{D}_s(\mathbf z_s, \overline{\mathbf W}) .
\end{aligned}
\end{equation}
The conditioning signal $\overline{\mathbf W}$ is a full-resolution $\mathbf W$-channel waveform provided to the spatial decoder $\mathcal{D}_s$. During training, $\overline{\mathbf W} = \mathbf W$, while at inference, $\overline{\mathbf W}=\widetilde {\mathbf W}$. Inspired by SpatialCodec~\cite{spatialcodec}, we condition the spatial decoder on ground-truth $\mathbf W$ during training to avoid tying the adapter to the reconstruction characteristics of a specific pretrained mono codec. This design facilitates its compatibility with different mono codecs. 

\ours{} uses two residual vector quantization (RVQ) codebooks, each containing 2,048 entries and operating at 12.5 tokens/s, yielding an additional bitrate of $R_s=2 \times 12.5 \times \log_2(2048) = 275~\text{bit/s}$. The total bitrate is therefore $R_m+R_s$, where $R_m$ is determined by the pretrained monaural codec. Although the fixed source--receiver geometry motivates a lower spatial token rate, preliminary experiments with lower rates degraded spatial reconstruction.

\subsection{Spatial decoder}

The spatial decoder $\mathcal{D}_s$ reconstructs all four FOA channels by modulating features of the conditioning waveform $\overline{\mathbf W}$ with the quantized spatial representation $\mathbf z_s$.
It consists of a waveform feature extractor $\mathcal{F}_W$, a spatial upsampling and modulation network $\mathcal{U}_s$, and an output network $\mathcal{G}_s$.
$\mathcal{F}_W$ extracts features from the reference waveform $\overline{\mathbf W}$, while $\mathcal{U}_s$ upsamples the quantized spatial representation to generate scale and bias parameters:
\begin{equation}
\mathbf h_W = \mathcal{F}_W(\overline{\mathbf W}), \quad
(\boldsymbol{\gamma},\boldsymbol{\beta})
= \mathcal{U}_s(\mathbf z_s).
\end{equation}
We use a variant of feature-wise linear modulation (FiLM)~\cite{film} to condition the waveform features on the spatial representation:
\begin{equation}
\label{eq:conditioning}
\widetilde{\mathbf h}_W
= \mathbf h_W \odot
\bigl(1+0.5\tanh(\boldsymbol{\gamma})\bigr)
+ \boldsymbol{\beta}.
\end{equation}
where $\odot$ denotes element-wise multiplication and the bounded scale limits excessive feature rescaling.
The output network $\mathcal{G}_s$ then reconstructs the complete FOA signal:
\begin{equation}
\widehat{\mathbf S}
= \mathcal{G}_s(\widetilde{\mathbf h}_W)
= \mathcal{D}_s(\mathbf z_s,\overline{\mathbf W}).
\end{equation}

\subsection{Spatial representation learning}
\label{sec:spatial_representation}

To encourage spatial tokens to capture spatial information, we use direct spatial supervision and spatial contrastive learning. 

\noindent \textbf{Spatial supervision.} 
An auxiliary head predicts intensity direction and log intensity
magnitude from spatial tokens, using targets derived from ground-truth FOA. The direction,
magnitude, and temporal smoothness losses form $\mathcal{L}_{\rm aux}$.

\noindent \textbf{Spatial contrastive learning.} 
We encourage spatial token representations to be similar for the same source direction and distinct for different directions.
Ideally, a positive pair consists of different utterances recorded under the same spatial condition. Since such pairs are not always available in spatial audio datasets, we approximate them using two temporal segments from the same recording with fixed source and receiver positions. Each 4-s FOA crop is split into two 2-s segments, which are independently encoded and quantized. Temporal mean pooling followed by $\ell_2$ normalization of the quantized embeddings produces the two representations forming a positive pair. Negatives include both segments of other batch recordings whose metadata source directions differ from the anchor's by more than $5^\circ$ after rotation augmentation. Alternative spatial renderings of the same utterance are included as hard-negative candidates, subject to the same angular criterion. We compute InfoNCE~\cite{oord2018cpc} using cosine similarity, averaging over all positive pairs with each segment serving in turn as the anchor to obtain $\mathcal{L}_{\mathrm{con}}$. 

\subsection{Training objective}

We supervise all four reconstructed FOA channels using waveform
$L_1$ and multi-resolution STFT losses, collectively denoted
by $\mathcal{L}_{\mathrm rec}$. To preserve spatial relationships between the omnidirectional and directional components, we apply diffuseness-weighted spatial-consistency loss ($\mathcal{L}_{\mathrm{sc}}$) introduced in FOA Tokenizer~\cite{foatokenizer}.

We employ multi-scale and multi-period discriminators following
HiFi-GAN~\cite{hifigan}. The channel-wise discriminators share parameters across the four channels and evaluate each channel independently, while the
joint discriminators distinguish ground-truth FOA signals $\mathbf{S}$
from reconstructions $\widehat{\mathbf{S}}$.
Both discriminator branches provide hinge adversarial
($\mathcal{L}_{\rm adv}$) and normalized feature-matching
($\mathcal{L}_{\rm fm}$) losses.

The full objective for \ours{} is a weighted sum of seven terms:
\begin{equation}
\nonumber
\mathcal L_G\!=\!\lambda_1\mathcal L_{\mathrm{rec}}
\!+\!\lambda_2\mathcal L_{\mathrm{aux}}
\!+\!\lambda_3\mathcal L_{\mathrm{sc}}
\!+\!\lambda_4\mathcal L_{\mathrm{con}}
\!+\!\lambda_5\mathcal L_{\mathrm{adv}}
\!+\!\lambda_6\mathcal L_{\mathrm{fm}}
\!+\!\lambda_7\mathcal L_{\mathrm{vq}} .
\end{equation}
where $\mathcal{L}_{\rm aux}$ and $\mathcal{L}_{\rm con}$
are defined in Section~\ref{sec:spatial_representation},
and $\mathcal{L}_{\rm vq}$ combines codebook and commitment losses.

%% file: sec/3_experimental_setup.tex
\section{Experiments}
\label{sec:exp}
\subsection{Experimental Setup}
\input{tables/main_table}

\noindent \textbf{Data and augmentation.}
We conducted experiments on the Spatial LibriSpeech dataset~\cite{spatiallibrispeech}. We reserved 5\% of the training rooms for validation, leaving 163,345 recordings from 8,504 rooms (544.53 hours) for training.
Evaluation used the official test set of 49,505 spatialized recordings from 73 speakers across 4,970 simulated rooms (85.49 hours).
During training, we randomly sampled token-aligned 4-s segments and padded shorter recordings.
Audio was resampled from 16 to 24 kHz to match the sampling rate of the pretrained mono codecs. For data augmentation, we applied a shared gain sampled uniformly from $[-6,6]$ dB to all four channels and random 3D rotations, sampled as normalized Gaussian quaternions, jointly to the directional channels and metadata DOAs.

\noindent \textbf{Implementation details.}
The spatial encoder and decoder use five causal convolutional stages with sampling factors $[8,6,5,4,2]$. The latent dimension is 384, and the spatial branch contains 28.84M parameters. We adopted the improved RVQ design from DAC~\cite{dac} to improve codebook utilization. We froze the pretrained monaural tokenizer and trained the spatial branch from scratch for 250k generator updates with a batch size of 8, using AdamW with a peak learning rate of $10^{-4}$ and cosine decay. Adversarial training began after 50k updates, with adversarial and feature-matching weights linearly increased over the next 10k updates.

\subsection{Baselines and Evaluation Metrics}

\noindent \textbf{Baselines.}
We compared the \ours{} framework with channel-independent (CI) coding using DAC-1, 2, 8~\cite{dac}, Mimi-8~\cite{moshi}, and MOSS-3, 8~\cite{mosstokenizer}, where suffixes indicate codebook counts, and Opus mapping family 2~\cite{opus} at 24 kbps total. The same adapter checkpoint was used across all mono codecs without retraining. MPEG-H 3D Audio~\cite{herre2014mpegh} operating in HOA coding at 128 kbps serves as a high-bitrate reference. FOA Tokenizer~\cite{foatokenizer} was excluded from the comparison because its source code and pretrained model are not publicly available.

\noindent \textbf{Objective evaluation.}
Following SpatialCodec~\cite{spatialcodec}, we evaluated spatial fidelity using relative transfer function (RTF) error in radians and MUSIC-based direction-of-arrival (DOA) error~\cite{MUSIC} (azimuth, elevation, and overall angle) in degrees, comparing estimates obtained from original and reconstructed signals.
We evaluated speech quality and intelligibility on monaural signals obtained by beamforming each original--reconstructed FOA pair toward the same active-intensity direction estimated from the original signal,
using non-intrusive UTMOS~\cite{UTMOS} and reference-based PESQ, SNR, and STOI. We also measured word error rate (WER) using Whisper-large-v3~\cite{whisper}.
Metrics are averaged over recordings, except corpus WER,
which pools word-error and reference-word counts.

\noindent \textbf{Subjective evaluation.}
We conducted a MUSHRA-inspired listening study with 27 participants using eight excerpts randomly selected from the Spatial LibriSpeech test-clean split. Participants rated overall speech quality considering speech naturalness and consistency of the perceived source direction with the reference, under six conditions: 1) MOSS-8 with \ours{}, 2) MOSS-8 CI, 3) MOSS-3 CI at a comparable total bitrate of 1.5 kbps, 4) Opus mapping family 2 (24 kbps total), 5) ground truth, and 6) a low-quality anchor. All listening-test samples were binaurally rendered using head-related transfer functions from~\cite{hrtf}. The anchor was generated by 3.5-kHz low-pass filtering the rendered ground truth and alternately hard-panning it left and right every 0.5~s.
Scores were averaged across excerpts for each participant and
system. Four codecs were compared using two-sided
paired $t$-tests ($df=26$), with Holm correction across all six pairwise comparisons ($\alpha=0.05$).

\begin{figure}[t]
    \centering
    \includegraphics[width=\linewidth]
        {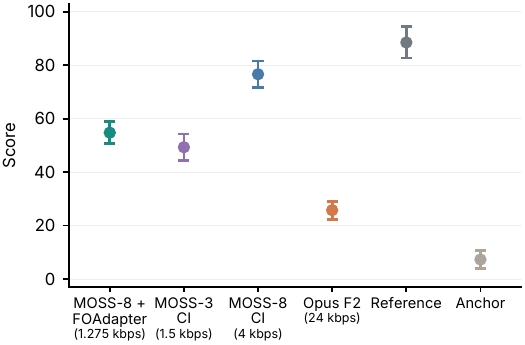}
    \caption{Subjective listening test results. Error bars indicate 95\% confidence intervals.}
    \label{fig:listening_test}
\end{figure}

\subsection{Spatial token swapping}
To examine the functional separation of speech content and source direction, we randomly selected 1000 recordings from the Spatial LibriSpeech test set and formed 500 different-speaker pairs using random 8-s crops. For each pair $(i,j)$, we generated
\[
\widehat{\mathbf{S}}_{i\leftarrow j}
= D_s\!\left(\mathbf{z}_{s,j},\widetilde{\mathbf{W}}_i\right),
\]
where $\widetilde{\mathbf{W}}_i$ denotes the MOSS-8 reconstruction of recording $i$'s $\mathbf{W}$ channel and $\mathbf{z}_{s,j}$ contains spatial tokens from recording $j$. Swapping in both directions yielded 1,000 cases.

Let $d_i$ and $d_j$ denote the angular distances between the estimated output direction and those of recordings $i$ and $j$. We define directional proximity to the spatial-token donor as $P_{\mathrm{dir}} = 100 d_i/(d_i+d_j)$, where a higher value indicates greater proximity to the spatial-token donor. We additionally report the percentage of cases with $d_j<d_i$. To evaluate content preservation, we beamformed each output toward the direction of $j$ and computed WER against ASR transcriptions of the original crops from both $i$ and $j$.

%% file: tables/main_table.tex
\begin{table*}[t]
\centering
\caption{Reconstruction on the Spatial LibriSpeech test set.
\ours{} framework bitrates include a 0.275-kbps spatial stream.}
\label{tab:main_table}
\small
\setlength{\tabcolsep}{4.5pt}
\renewcommand{\arraystretch}{0.95}
\setlength{\aboverulesep}{0.3ex}
\setlength{\belowrulesep}{0.3ex}
\resizebox{\textwidth}{!}{%
\begin{tabular}{@{}llc@{\hspace{10pt}}cccc@{\hspace{10pt}}cccc@{\hspace{10pt}}c@{}}
\toprule
& & Total bitrate
& \multicolumn{4}{c}{Spatial fidelity}
& \multicolumn{4}{c}{Speech quality}
& Intelligibility \\
\cmidrule(lr){3-3}
\cmidrule(lr){4-7}
\cmidrule(lr){8-11}
\cmidrule(lr){12-12}
Codec & Coding & (kbps)
& Az.$\downarrow$ & Ele.$\downarrow$
& DOA$\downarrow$ & RTF$\downarrow$
& UTMOS$\uparrow$ & PESQ$\uparrow$
& SNR$\uparrow$ & STOI$\uparrow$
& WER$\downarrow$ \\
\midrule
\multicolumn{2}{@{}l}{Ground Truth} & --
& -- & -- & -- & --
& 2.127 & 4.644 & $\infty$ & 1.000 & 4.47 \\
\midrule
DAC-1 & CI & 3.000
& 12.95 & 6.65 & 15.69 & 1.5271
& 1.277 & 1.207 & 1.31 & 0.716 & 51.95 \\
& + \ours{} & 1.025
& 4.45 & 4.26 & 6.86 & 1.4893
& 1.269 & 1.146 & $-0.19$ & 0.656 & 76.10 \\
\addlinespace[1pt]
DAC-2 & CI & 6.000
& 4.70 & 3.39 & 6.42 & 1.521
& 1.369 & 1.560 & 3.34 & 0.802 & 23.22 \\
& + \ours{} & 1.775
& 4.43 & 4.18 & 6.78 & 1.4879
& 1.319 & 1.338 & 1.23 & 0.732 & 42.03 \\
\addlinespace[1pt]
DAC-8 & CI & 24.000
& 1.31 & 0.98 & 1.93 & 1.467
& 1.907 & 3.429 & 9.37 & 0.938 & 6.43 \\
& + \ours{} & 6.275
& 4.41 & 4.15 & 6.74 & 1.4877
& 1.581 & 2.118 & 4.41 & 0.849 & 10.52 \\
\addlinespace[1pt]
Mimi-8 & CI & 4.400
& 11.31 & 4.82 & 13.04 & 1.5303
& 1.812 & 1.815 & 3.94 & 0.802 & 16.15 \\
& + \ours{} & 1.375
& 4.36 & 4.10 & 6.66 & 1.4938
& 1.740 & 1.614 & 1.89 & 0.773 & 21.53 \\
\addlinespace[1pt]
MOSS-8 & CI & 4.000
& 5.74 & 3.84 & 7.65 & 1.5220
& 2.208 & 2.107 & 4.79 & 0.844 & 7.34 \\
& + \ours{} & 1.275
& 4.35 & 4.14 & 6.68 & 1.4942
& 1.969 & 1.845 & 2.46 & 0.813 & 9.13 \\
MOSS-3 & CI & 1.500
& 21.72 & 8.21 & 24.52 & 1.5325
& 1.838 & 1.482 & 1.55 & 0.744 & 13.79 \\
\midrule
\multicolumn{2}{@{}l}{MPEG-H 3D Audio (HOA)} & 128.000
& 0.16 & 0.11 & 0.26 & 1.158
& 2.039 & 4.409 & 22.16 & 0.992 & 4.64 \\
Opus-F2 &   & 24.000
& 3.43 & 1.93 & 4.30 & 1.4926
& 1.268 & 1.692 & 5.22 & 0.869 & 10.69 \\
\bottomrule
\end{tabular}%
}
\end{table*}

%% file: sec/4_experimental_results.tex
\section{Results}
\label{sec:results}


\subsection{Reconstruction performance}

Table~\ref{tab:main_table} summarizes the reconstruction results, highlighting the effectiveness of combining a pretrained mono codec with a compact spatial stream at low total bitrates. With a fixed 0.275~kbps of spatial overhead, the \ours{} framework achieved consistent DOA errors of 6.66--6.86$^\circ$ across all evaluated backbones while reducing the total bitrate by 66\%--74\% relative to the corresponding channel-independent (CI) systems. Speech quality varied across backbone monaural codecs. At comparable bitrates, MOSS-8 with \ours{} (1.275~kbps) outperformed MOSS-3 CI (1.500~kbps) across all speech-quality metrics, reducing WER from 13.79\% to 9.13\% and DOA error from 24.52$^\circ$ to 6.68$^\circ$. Higher-rate CI codecs and MPEG-H nevertheless provided better reconstruction quality or localization accuracy.

\subsection{Spatial token swapping}
Spatial token swapping supports a functional separation between speech content and source direction information. The output direction largely follows the spatial tokens from recording $j$, while recognized speech content follows the reference waveform from recording $i$. Across 1,000 swap cases, the mean directional proximity score was 88.32\%, and the output was closer to the direction of $j$ in 97.4\% of cases. The mean angular distances to $j$ and $i$ were $8.48^\circ$ and $91.12^\circ$, respectively. WER was 9.55\% against the ASR transcription of $i$, compared with 106.77\% against that of $j$. These findings suggest that spatial tokens can serve as a transferable control signal for source direction across different speakers.


\subsection{Subjective evaluation results}


Subjective listening test results support the perceptual bitrate efficiency of \ours{}. Figure~\ref{fig:listening_test} shows that MOSS-8 with \ours{} scored significantly higher than MOSS-3 CI (54.78 vs.\ 49.33; $p=0.026$) at a 15\% lower total bitrate. It also outperformed 24-kbps Opus-F2 (25.78; $p<0.001$), but scored lower than MOSS-8 CI (76.63; $p<0.001$). The lower subjective score of Opus-F2, despite its favorable DOA, SNR, and STOI results, suggests that these metrics alone may not fully capture overall perceptual quality.

%% file: sec/5_conclusion.tex
\section{Conclusion}
\label{sec:conclusion}


We present \ours{}, a plug-and-play spatial adapter that extends pretrained mono codecs to FOA speech coding using a 275-bit/s spatial stream. Across multiple mono codecs, \ours{} provides consistent directional reconstruction at substantially lower bitrates than CI coding, while spatial-token swapping indicates partial separation of speech and spatial information. The resulting compact spatial tokens offer a promising representation for spatial speech generation.
